\documentclass[aps,showpacs,twocolumn,superscriptaddress]{revtex4}
\usepackage{graphicx}
\usepackage{dcolumn}
\def\tstrut{\vrule height2.5ex depth0pt width0pt} 
\begin{document}

 \title {$K^-$-Nucleus Scattering at Low and Intermediate Energies.}
\author{C. Garc\'\i a-Recio}
\author{A.J. Melgarejo}  
\author{J.Nieves}
\affiliation{Departamento de F\'{\i}sica Moderna, Universidad de Granada, 
 E-18071 Granada, Spain}

 \begin{abstract} \rule{0ex}{3ex} We calculate $K^-$-nucleus elastic
 differential, reaction and total cross sections for different nuclei
 ($^{12}$C,$^{40}$Ca and $^{208}$Pb ) at several laboratory antikaon
 momenta, ranging from $127$ MeV to $800$ MeV. We use different
 antikaon-nucleus optical potentials, some of them fitted to kaonic
 atom data, and study the sensitivity of the cross sections to the
 considered antikaon-nucleus dynamics.

\end{abstract}

\maketitle

 \section{Introduction}

 The works on meson baryon dynamics of
 Ref.~\cite{KSW95} showed how Chiral symmetry constraints  could
 be accommodated within a unitarity approach, able to describe
 resonances. This proved to be crucial
 to disentangle the intricate interaction between antikaons and
 nucleons at low energies~\cite{RO97}--\cite{Ga02}. The model of
 Ref.~\cite{RO97}, was employed in Ref.~\cite{RO00} to microscopically
 derive an optical potential for the $K^-$ in nuclear matter in
 a self-consistent manner.   Self-consistency
 turned out to be a crucial ingredient to derive the $K^- $-nucleus
 potential and led to an optical potential considerably more shallow
 than those found in Refs.~\cite{BF97}--\cite{FG99}. 

 In Refs.~\cite{zaki} and~\cite{baca}, the predictions of the chirally
 inspired potential of Ref.~\cite{RO00} for measured shifts
 and widths of $K^-$ atoms were evaluated, and it was found that this
 potential provides an acceptable description of the observed kaonic
 atom states, through the whole periodic table. Despite of having both
 real and imaginary parts of quite different depth, some other
 empirical optical potentials (\cite{FG93}--\cite{FG99}) also examined
 in Ref.~\cite{baca}, led to acceptable descriptions of the
 experimentally available $K^-$ atom data as well. However, there were
 appreciable differences among the predicted widths for deeply bound
 antikaon nuclear states, not detected yet, when different potentials
 were used.   The aim of this
 paper is to explore the possibility of differentiating between
 several $K^-$ nucleus optical potentials by means of the
 scattering data. The extrapolation to finite $K^-$ kinetic energies
 of the potential of Ref.~\cite{RO00} requires at least the inclusion
 of the $p$-wave part of the $K^-$ selfenergy. This was performed
 in Ref.~\cite{pwave}, and tested for $K^- p$ scattering in
 Ref.~\cite{Ji02}.  However, even after having included $p$-wave
 contributions, one cannot expect reliable predictions from the
 theoretical potential of Refs.~\cite{RO00} and~\cite{pwave}, at the
 lowest energy for which there exist experimental data ($800$
 MeV for the $K^-$ momentum), where $d$ and $f$ waves contributions
 are relevant. Besides, as
 we will show, for this relatively high energy, the impulse
 approximation works reasonably well, which is a clear indication that
 these data do not have much information on the details of the
 $K^-$-nucleus dynamics. Thus, we have also focused our attention at
 the typical momentum of the $K^-$ after the $\phi$-meson decay
 ($\approx$ 127 MeV) with the hope that the scattering experience
 could be performed at DA$\Phi$NE or at KEK or in the future Japanese
 Hadron Collider (JHC).

\section{$K^-$-Nucleus Optical
Potentials}
We solve the Klein Gordon equation
\begin{equation}
\left ( -\vec{\nabla}^2 + \mu^2 + 2\omega V_{\rm opt}\right)\Psi = 
\left ( \omega - V_C \right )^2 \Psi, \label{eq:kge}
\end{equation}
where $\omega$ is the Center of Mass (CM) $K^-$--nucleus 
energy, $V_C$ and $V_{\rm opt}$ are the finite-size
Coulomb and optical $K^-$--nucleus potentials and 
$\mu$ the reduced $K^-$-nucleus mass. At large distances and
for a CM scattering angle $\theta$, the
$K^-$ wave-function $\Psi(\vec{r})$ behaves as $ I(r) + f(\theta)
S(r)$, with $I(r)$ and $S(r)$ the standard wave functions for
Coulomb scattering from a punctual charge $Z$ and $f(\theta)$ the
scattering amplitude, which normalization is determined by its relation
to the CM differential elastic cross section $d\sigma_{\rm e}/ d\Omega
= \left | f(\theta) \right |^2$.  The integrated cross sections read:
\begin{eqnarray}
\sigma_{\rm e} &=& \frac{\pi}{q^2} \sum_l (2l+1) \left |1-
\eta_l e^{2{\rm i~} \left( \sigma_l+\delta_l\right)}\right |^2, \label{eq:se}
\\[2mm]
\sigma_{\rm t} &=& \frac{2\pi}{q^2} \sum_l (2l+1) \left [1-
\eta_l \cos\left(2 \left( \sigma_l+\delta_l\right)\right) \right
],\label{eq:sr}\\[2mm]
\sigma_{\rm re}&=& \frac{\pi}{q^2}\sum_l (2l+1) \left [1-
\eta_l^2\right] \label{eq:st}
\end{eqnarray}
with $q$ the CM $K^-$ momentum and $\sigma_l$, $\delta_l$ and $\eta_l$
the standard Coulomb phase shifts, the additional phase shifts due
to strong interaction and the inelasticities appearing in the standard
partial wave decomposition of $f(\theta)$ (see 
Ref.~\cite{NOG93}). While the elastic ($\sigma_{\rm e}$) and total
($\sigma_{\rm t}$) cross sections are infinite, the reaction
($\sigma_{\rm re}$) cross section is finite because of the short-range
of the nuclear interaction.

 The $K^-$-nucleus optical potential, $V_{\rm opt}$, is related to the
 $K^-$-selfenergy, $\Pi(q^0,|\vec q~|),$ inside of a nuclear medium, 
 neglecting isovector effects, by
\begin{equation} \label{eq:V1} 2 \omega V_{\rm
 opt}(r)=  \Pi\left(m+T,~ ({q^0}^2-m^2)^\frac12;~
 \rho(r)\right ) \end{equation} 
 where $m$ and $T$ are the $K^-$ mass and laboratory
 kinetic energy and $\rho$ is the sum of proton and
 neutron densities. 

 From the antikaon-selfenergy, as determined by, Refs.~\cite{RO00}
 and~\cite{pwave}, we define the first selfenergy used in
 this work ($\Pi^{\rm TH}$).  This selfenergy does not have any free
 parameters, all the needed input is fixed either from studies of
 meson-baryon scattering in the vacuum or from previous studies of
 pion-nucleus dynamics~\cite{NOG93}.  It provides an acceptable
 ($\chi^2/dof$  of 2.9) description of the set of 63
 shifts and widths of $K^-$ atom levels used in
 Ref.~\cite{baca}. We have neglected all type of non-localities, since
 they lead to changes in the results presented here of 3\% at most.

 As in Ref.~\cite{baca}, we also construct a
 modified selfenergy, which we call $ \Pi^{\rm THPH}$,
 by adding to $ \Pi^{\rm TH}$ a phenomenological part linear
 in density. This phenomenological 
 part is determined by a  constant
 $\delta b_0$ which we fix to the value 
$(0.12 - {\rm i~} 0.38 ) {\rm fm}$, obtained in Ref.~\cite{baca} 
 from a $\chi^2$-fit to the kaonic atom data. The new selfenergy reads: 
\begin{eqnarray} 
\label{eq:V1m} 
\Pi^{\rm THPH}(r) & = & \Pi^{\rm TH}(r)
 -4\pi~ \delta b_0 ~\rho(r) 
\end{eqnarray}
The third selfenergy considered in this work is just obtained from the
Impulse Approximation (IA), i.e., $t\rho$ form for the $K^-$ 
selfenergy, and it neglects all orders higher
than the leading one, in the density expansion. It reads:
\begin{eqnarray}
 \Pi^{\rm IA} (r) &=& -4\pi \frac{\sqrt{s}}{M}
~ b_0^{\rm IA} (\sqrt{s}~) \rho (r)   \label{eq:ia}
\end{eqnarray}
with $b_0^{\rm IA} (\sqrt{s}~) = \frac14 \left ( 3~ {_1f}(\sqrt{s}~) +~
{_0f}(\sqrt{s}~) \right )$, $M$ the nucleon mass,
$\sqrt{s}$ the total CM $K^- N$  energy and $_{I=1,0}f$ the
isoscalar and isovector forward antikaon-nucleon scattering
amplitudes, which partial wave decomposition reads:
\begin{equation}
_I f (\sqrt{s}~)  =  \sum_l \left ( (l+1)~ _I f_l^{j_+}(\sqrt{s}~)
+ l~ _I f_l^{j_-}(\sqrt{s}~)  \right ) \label{eq:deff}
\end{equation}
with $j_{\pm}=l\pm 1/2$ the total angular momentum. At threshold,
$b_0^{\rm IA_{thr}}\equiv b_0^{\rm IA} (m+M)= (-0.15+{\rm i~ }0.62)~
{\rm fm}$~\cite{Ma81}.  The IA leads to extremely poor results for
kaonic atoms~\cite{BF97},\cite{baca}. This is a clear indication that
higher density corrections, not taken into account within the IA, are
extremely important for kaonic atoms.

Finally, we have also considered two  other antikaon 
selfenergies fitted to the kaonic atom data and energy
independent (\cite{FG93} and~\cite{FG99}):
\begin{eqnarray}
\Pi^{\rm 2DD}  & = & - 4 \pi \left ( 1 +
\frac{\mu}{M}\right )   \left ( b_0^{\rm IA_{thr}} + B_0 \left (
 \frac{\rho}{\rho_0} \right )^\alpha \right ) \rho \label{eq:2DD} \\[2mm]
\Pi^{\rm IAPH}  & = & - 4 \pi \left ( 1 +
\frac{\mu}{M}\right ) {\tilde b}_0 \rho  \label{eq:IAPH} 
\end{eqnarray}
with ${\tilde b}_0 =(0.52+{\rm i }~ 0.80)~ {\rm fm}$,  
$B_0 = (1.62-{\rm i }~ 0.028) ~ {\rm fm}$ and $\alpha = 0.273$ as
determined from $\chi^2$-fits to $K^-$-atom data in Ref.~\cite{baca}.
Note that, though both $\Pi^{\rm IAPH}$ and $\Pi^{\rm IA}$ are linear
in density selfenergies, they lead to substantially different
potentials,  since the real parts
of the coefficients ${\tilde b}_0$ and $b_0^{\rm IA_{thr}}$ differ
both in sign and in size.

\section{Results and Concluding Remarks}
\label{sec:concl}

Since the $K^-$ lifetime is relatively small, in practical terms it is
experimentally difficult to count with low energetic $K^-$
beams. However, all selfenergies described in the previous section,
except for that obtained in the IA, are valid only near
threshold. Thus, we have studied the case $q_{\rm lab} = 127$ MeV,
since this is the $K^-$ momentum after the $\phi$-meson decay. 
\begin{figure}
\centerline{\includegraphics[height=14.8cm]{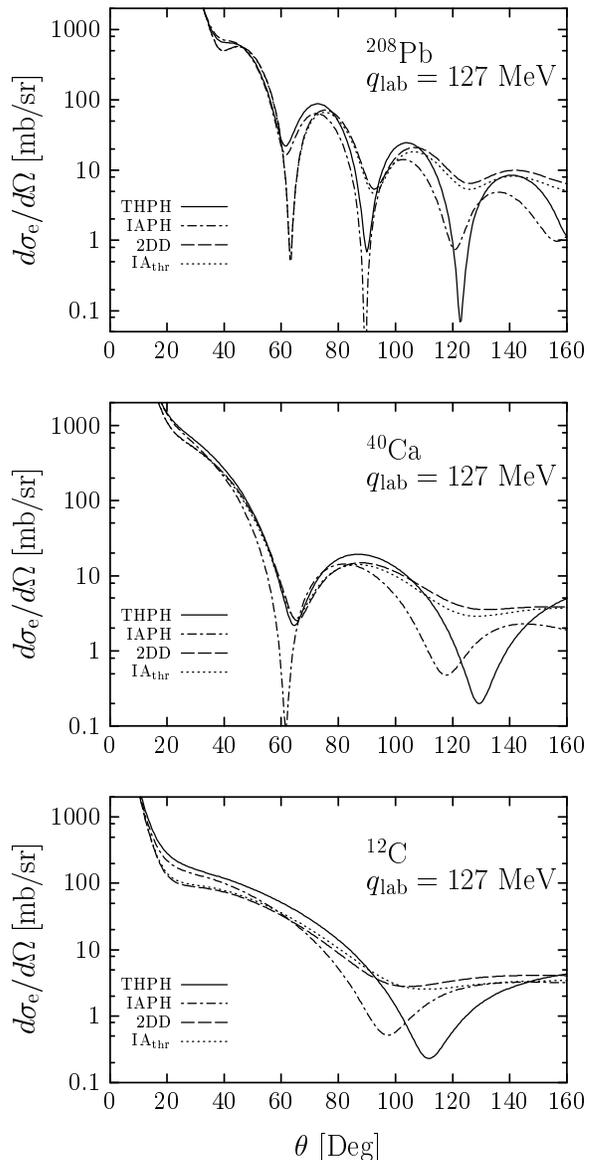}}
\caption[pepe]{\footnotesize CM cross sections for elastic
scattering of $q_{\rm lab}=127$ MeV $K^-$ from $^{12}$C, $^{40}$Ca and
$^{208}$Pb with different $K^-$ selfenergies.}
\label{fig:comp1}
\end{figure}
In Fig.~\ref{fig:comp1} we present results obtained with the $K^-$
selfenergies fitted to the kaonic atom data.  We also show results
obtained by using the IA, where we have approximated the IA selfenergy
at $q_{\rm lab}=127$ MeV by its threshold value quoted above. Strong
interaction integrated elastic, reaction and total cross sections are
also given in the top part of Table~\ref{tab:comp1}. We obtain these
cross sections after having got rid of the Coulombian interaction,
otherwise the total and elastic cross sections would diverge, i.e., we
compute strong phase-shifts and inelasticities ($\delta_l$ and
$\eta_l$) in presence of the Coulomb interaction, and afterwards we
set to zero the Coulombian phase shifts, $\sigma_l$, in the formulae
of Eqs.~(\ref{eq:se})-(\ref{eq:st}).
\begin{figure}
\centerline{\includegraphics[height=9.6cm]{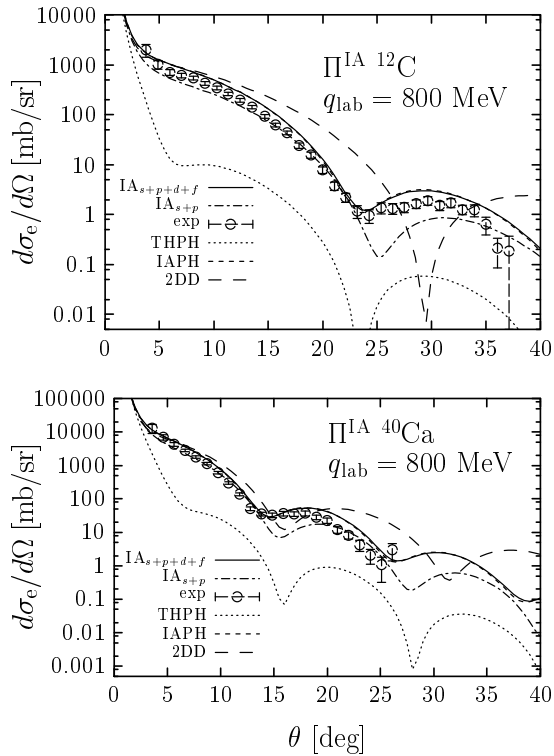}}
\caption {\footnotesize CM differential cross section for elastic
scattering of $q_{\rm lab}=800$ MeV $K^-$ from $^{12}$C and $^{40}$Ca.
Data are taken from Ref.~\protect\cite{Ma82}.}
\label{fig:800}
\end{figure}
\begin{table}
\begin{center}
\begin{tabular}{l|ccc|ccc|ccc}\hline
& \multicolumn{3}{c|}{$^{12}$C} & \multicolumn{3}{c|}{$^{40}$Ca}&
\multicolumn{3}{c}{$^{208}$Pb}\\\tstrut 
Potential & $\sigma_{\rm e}$ & $\sigma_{\rm re}$ & $\sigma_{\rm
t}$ & $\sigma_{\rm e}$ & $\sigma_{\rm re}$ & $\sigma_{\rm
t}$ & $\sigma_{\rm e}$ & $\sigma_{\rm re}$ & $\sigma_{\rm
t}$ \\\hline\tstrut
$\Pi^{\rm THPH}$ & 444 & 501 & 945 & 1190 & 1201 & 2391 &4345
& 4219 & 8564 \\\tstrut
$\Pi^{\rm 2DD}$ & 370 & 415 & 785 & 1064 & 1024 & 2088 &4287
& 3667 & 7954 \\\tstrut
$\Pi^{\rm IAPH}$ & 411 & 568 & 979 & 1029 & 1313 & 2342 &4094
& 4363 & 8457 \\\tstrut
$\Pi^{\rm
IA}|_{\protect\sqrt{s~}= m+M} $
& 380 & 420 & 800 & 1040 & 1043 & 2083 &4264
& 3699 & 7963  \tstrut \\      \hline\hline
\tstrut
$\Pi^{\rm THPH}$ & 71  & 250 & 321 &  242 &  572 &  814 &1329
& 2074 & 3403 \\\tstrut
$\Pi^{\rm 2DD}$ & 252 & 312 & 564 & 545  &  737 & 1282 &2057
& 2288 & 4345 \\\tstrut
$\Pi^{\rm IAPH}$ & 250 & 374 & 624 & 613  & 858  & 1471 &2115 
& 2562 & 4677 \\\tstrut
$\Pi^{\rm IA}|_{s+p} $
& 248 & 384 & 632 & 615  & 874  & 1489 &  2148  
& 2557     &  4705 \\\tstrut
$\Pi^{\rm
IA}|_{s+p+d} $
& 278 & 406 & 684 & 653  & 913  & 1566 &  2195  
& 2642     & 4837  
     \tstrut \\      \hline
\end{tabular}
\end{center}
\caption{ \footnotesize Strong integrated elastic, reaction and total 
cross sections (in mb) at $q_{\rm lab}=127$ (top) and 300 (bottom) MeV.}
 \label{tab:comp1}
 \end{table}
As can be seen in the figure, the non-linear density dependent $K^-$
selfenergy, $\Pi^{\rm 2DD}$, and the linear density dependent,
threshold IA selfenergy, $\Pi (\sqrt{s~}) = \Pi^{\rm
IA}|_{\protect\sqrt{s}= m+M}$, provide extraordinarily similar
results. Since both models have the same linear term in density, this
is a clear indication that the reaction takes place in the surface of
the nuclei, because of the big imaginary part of the potentials. The
semiphenomenological $\Pi^{\rm THPH}$ selfenergy, has a stronger
departure from a linear behaviour in density than $\Pi^{\rm 2DD}$, it
has a smaller imaginary part (see Fig. 1 of Ref.~\cite{baca}) and all
of these explain the bigger differences with the IA model.  Results,
in particular position of the minima, obtained with $\Pi^{\rm THPH}$
are clearly distinguishable from those obtained with any of the other
three models also plotted in the figure, pointing out to a clear
different density behavior likely due to the selfconsistent derivation
of it. As a matter of example, for $^{12}$C in the region around
$\theta=60^{\rm o}$, 2DD, IA and IAPH give similar elastic cross
sections of about 33 mb/sr, whereas THPH gives about 52 mb/sr. This
difference is appreciable and the size of the cross sections, tens of
mb/sr, might allow to measure such a difference at DA$\Phi$NE or KEK
or in the future at the JHC. The differences are even bigger for
larger angles, around the minimum of the THPH cross section (region
110-$130^{\rm o}$), but there, the cross sections are smaller, which
makes harder to get the required statistics to see the
effect. Besides, theoretical results in the neighborhood of a minimum
are subject to more uncertainties. Similar conclusions can be drawn
from the $^{40}$Ca and $^{208}$Pb results. In what respects to the
integrated cross sections of Table~\ref{tab:comp1}, 2DD and IA give
similar cross sections, though the IA reaction cross section is always
slightly bigger, because the imaginary part of the $B_0$ parameter in
Eq.~(\ref{eq:2DD}) is negative. The IAPH model always provides the
biggest reaction cross sections, because its selfenergy has also the
largest imaginary part among all models considered (see Fig. 1 of
Ref.~\cite{baca}). Thus one can differentiate two sets of models, i.e,
2DD and IA selfenergies from THPH and IAPH ones. Besides,
measurements, with precisions of about 10\%, of the reaction cross
sections would disentangle between THPH and IAPH models.

Let us look now to the experimental data. There only exist
data~\cite{Ma82} on $K^-$ differential elastic cross sections for
$q_{\rm lab}=800$ MeV and from $^{12}$C and $^{40}$Ca. In
Fig.~\ref{fig:800} we compare the IA predictions (solid line),
including up to $f-$waves (from Ref.~\cite{Go77}), to data.  There
also exist some data on total cross sections (mb), $338 \pm 8$ [$306
\pm 8$],~\cite{Bu68} from $^{12}$C at $q_{\rm lab}=800$ [$655$]
MeV. The IA$_{s+p+d+f}$ model provides again an acceptable
description: 345 mb at 800 MeV and 304 mb at 655 MeV.  Thus, this
region is less sensitive to in nuclear medium effects than the $K^-$
atom one.  Indeed, the Glauber approximation also describes the 800
MeV scattering data, as discussed in~\cite{SC99}. This work also
corroborates that the imaginary part of the $K^-N$ amplitude, obtained
from the $K^-$-nucleus scattering data, is close to that deduced in
the vacuum. Besides, the contribution of $d$ and $f$ waves, not
included in the THPH model, turn out
to be important (compare the solid line to the IA results obtained
when only $s$ and $p$ waves --dot-dashed line-- are considered, in
Fig.~\ref{fig:800}).  In addition, the models of Refs.~\cite{RO97} and
~\cite{pwave} for the $s$ and $p$ $K^-N$ waves, though realistic near
threshold, can not be safely extrapolated to momenta as high as 800
MeV. Thus, one expects the poor description of data provided by
the THPH model. It is however surprising, that the IAPH predictions
turn out to be almost indistinguishable from the IA$_{s+p+d+f}$
ones. This is merely a coincidence  and, it occurs since
accidentally at this momentum, $b_0^{\rm IA} (\sqrt{s}) $ is
approximately equal to ${\tilde b}_0$.
\begin{figure}
\centerline{\includegraphics[height=14.7cm]{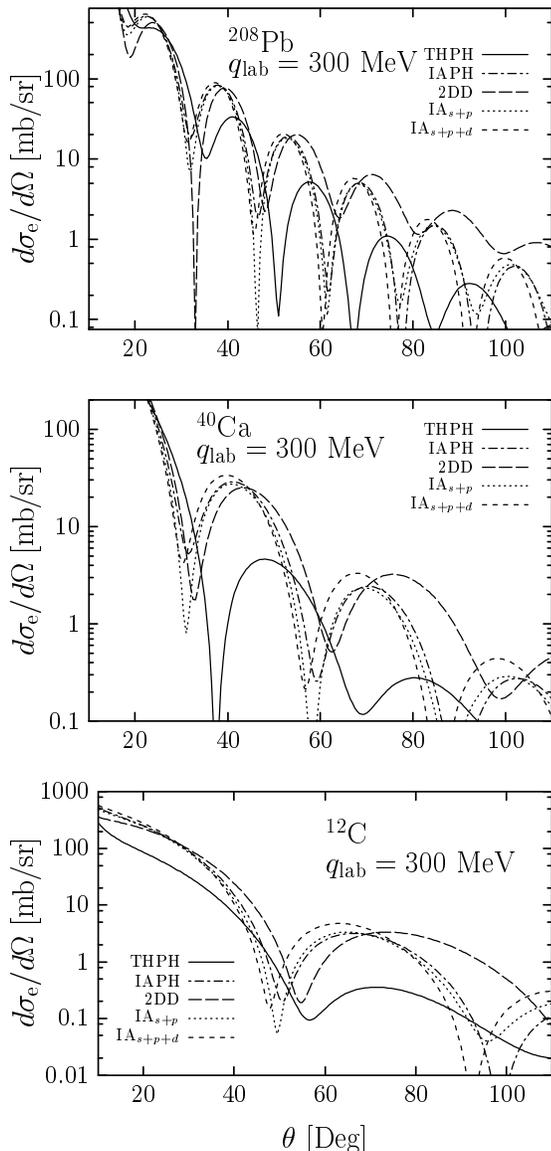}}
\caption {\footnotesize CM cross sections for elastic
scattering of $q_{\rm lab}=300$ MeV $K^-$ from $^{12}$C, $^{40}$Ca and
$^{208}$Pb with different $K^-$ selfenergies.}
\label{fig:comp2}
\end{figure}

To finish, we also present results at an intermediate $K^-$ momentum
($q_{\rm lab}=300$ MeV), despite the fact that there exist no
data. For this momentum, calculations based on the IA shows that
higher waves than the $p$ one have a small/moderate contribution and
therefore can be neglected in some approximations. Thus in
Fig.~\ref{fig:comp2} and bottom part of Table~\ref{tab:comp1}, we
compare again the THPH, 2DD and IAPH models for the $K^-$ selfenergy
inside the nuclear medium, together with the IA results including up
to the $p$-wave, or up to the $d$-wave (partial waves are taken from
Ref.~\cite{Go77}). The first observation is that the 2DD model differs
now more than for the $127$ MeV momentum case, from the IA
models. This is mainly due to the effect of $p$-wave in the latter
ones. The second observation is that the semiphenomenological model
THPH leads to a pattern clearly different than the rest of
selfenergies, not only for the elastic differential cross section but
also in the integrated ones compiled in Table~\ref{tab:comp1}. This is
in principle good news, because then a scattering measurement in this
region of $K^-$ momentum will be definitive to disentangle between
this approach and the others considered in Fig.~\ref{fig:comp2} and
bottom of Table~\ref{tab:comp1}. However a word of caution must be said here,
the $s$-wave part of the antikaon selfenergy of Ref.~\cite{RO00} is
based on a model for the $K^-N$ scattering in the free space that,
though it is quite successful near threshold, predicts amplitudes for
the isoscalar channel around $q_{\rm lab}=300$ MeV, with real parts
which are in total disagreement (in sign and in size, see
Ref.~\cite{ORB02}) with the analysis of Ref.~\cite{Go77}. Thus, most
probably one cannot trust the THPH model to describe the
$K^-$-dynamics at this momentum. Indeed, there is no reason either to
believe more in the 2DD and IAPH models, and we believe that the more
reliable predictions for $q_{\rm lab}= 300$ MeV are those based on the
IA.

\section*{Acknowledgments}
We thank to E.Oset and A.Ramos for useful
discussions. A.J.M. acknowledges a fellowship from the University of
Granada. This work was supported by  Junta de Andaluc\'\i a, DGI and
FEDER funds (BFM2002-03218).

\end{document}